\documentclass[12pt]{article}
\usepackage{a4wide}
\usepackage[utf8]{inputenc}
\usepackage{graphicx}
\usepackage{changepage}
\usepackage{tabularx}
\usepackage{graphics}
\usepackage{array}
\usepackage{amsmath}
\usepackage{amssymb}
\usepackage{epsfig,cite,scalefnt,graphicx, setspace,slashed,wrapfig,enumitem,amsthm,braket,mathrsfs,longtable}
\usepackage{subcaption}
\numberwithin{equation}{section}
\usepackage{tikz}
\DeclareSymbolFont{bbold}{U}{bbold}{m}{n}
\DeclareSymbolFontAlphabet{\mathbbold}{bbold}

\def\nn{\nonumber\\}
\def\msbar{\footnotesize{$\overline{\textsc{MS}}\;$}\normalsize}
\renewcommand*{\thefootnote}{\fnsymbol{footnote}}

\usetikzlibrary{intersections}
\usetikzlibrary{calc}
\usetikzlibrary{shapes}
\usetikzlibrary{positioning,arrows}
\usetikzlibrary{decorations.pathmorphing}
\usetikzlibrary{decorations.markings}
\usetikzlibrary{shapes.geometric}

\newcommand{\dd}{\mathop{}\!\mathrm{d}}
\newcommand{\one}{\scalebox{1.25}{$\mathbbold{1}$}}

\newcommand{\cofg}[2]{\mathfrak{g}^{(#1)}_{#2}\,}

\newcommand{\cofy}[2]{\mathfrak{y}^{(#1)}_{#2}\,}
\newcommand{\cofA}[2]{\mathcal{A}^{(#1)}_{#2}\,}
\newcommand{\cofT}[2]{\mathcal{T}^{(#1)}_{#2}\,}

\newcommand{\udindices}[2]{\phantom{}^{#1}\phantom{}_{#2}}

\usepackage[bookmarks=false]{hyperref}
\usepackage{xcolor}
\definecolor{rossoCP3}{cmyk}{0,.88,.77,.40}
\definecolor{blueRef}{rgb}{0.2,0.2,0.6}
\definecolor{blue}{rgb}{0,0.396,0.741}
\hypersetup{
	colorlinks,
	citecolor=blueRef, 		
	linkcolor=rossoCP3,	
	urlcolor=rossoCP3,			
}

\tikzset{
	chiral/.style={draw=black, postaction={decorate},
		decoration={markings,mark=at position .55 with {\arrow[black,scale=1.5]{stealth}}}},
	gauge/.style={decorate, draw=black,
		decoration={coil,aspect=0}},
	scalar/.style={draw=black, dashed}
}

\allowdisplaybreaks

\begin{document}
	
	\begin{titlepage}
		\begin{flushright}
			$\text{CP}^{3}$-Origins-2019-1 DNRF90\\
		\end{flushright}
		\date{}
		\vspace*{3mm}
		
	\begin{center}
		{\Huge Weyl Consistency Conditions and $\gamma_{5}$}\\[12mm]
		{\bf C.~Poole\footnote{\href{mailto:cpoole@cp3.sdu.dk}{\tt cpoole@cp3.sdu.dk}} and A.~E.~Thomsen\footnote{\href{mailto:aethomsen@cp3.sdu.dk}{\tt aethomsen@cp3.sdu.dk}}}\\

		\vspace{5mm}
		$\text{CP}^{3}$-Origins, University of Southern Denmark,\\ Campusvej 55, DK-5230 Odense M, Denmark\\
			
	\end{center}
		
	\vspace{3mm}
	\begin{abstract}
	The treatment of $\gamma_{5}$ in Dimensional Regularization leads to ambiguities in field-theoretic calculations, of which one example is the coefficient of a particular term in the four-loop gauge $\beta$-functions of the Standard Model. Using Weyl Consistency Conditions, we present a scheme-independent relation between the coefficient of this term and a corresponding term in the three-loop Yukawa $\beta$-functions, where a semi-na\"ive treatment of $\gamma_{5}$ is sufficient, thereby fixing this ambiguity. We briefly outline an argument by which the same method fixes similar ambiguities at higher orders.\\
	\end{abstract}
		
\vfill
		
\end{titlepage}

\renewcommand*{\thefootnote}{\arabic{footnote}}
\setcounter{footnote}{0}

\section{Introduction}
The treatment of $\gamma_{5}$ in Dimensional Regularization is a well-known theoretical issue \cite{TV72}, and can be summarized in the following statement: given a four-dimensional, Poincar\'e-invariant quantum field theory, there is no gauge-invariant regularization method that preserves chiral symmetry \cite{J00}. The precise connection is most easily demonstrated using the ABJ anomaly, the derivation of which requires
\begin{equation}
	\text{tr}\left[ \gamma^{\mu}\gamma^{\nu}\gamma^{\rho}\gamma^{\sigma}\gamma_{5} \right] = 4i\epsilon^{\mu\nu\rho\sigma}, \quad \epsilon_{0123} = - \epsilon^{0123} = 1
	\label{abj}
\end{equation}
in four dimensions, whereas the $d$-dimensional $\gamma$-matrix algebra
\begin{equation}
	\left\{ \gamma^{\mu}, \gamma^{\nu} \right\} = 2g^{\mu\nu}\one, \quad  g^{\mu\nu}g_{\mu\nu} = d, \quad \left\{ \gamma^{\mu}, \gamma_{5} \right\} = 0
\end{equation}
combined with trace-cyclicity directly implies
\begin{equation}
	\text{tr}\left[ \gamma^{\mu}\gamma^{\nu}\gamma^{\rho}\gamma^{\sigma}\gamma_{5} \right] = 0
\end{equation}
even when $d \rightarrow 4$. The correct trace-relation may be obtained by using the 't Hooft-Veltman algebra, which decomposes the $d$-dimensional space into a four-dimensional subspace containing $\gamma^{\mu} = \hat{\gamma}^{\mu}, \mu \in \{ 0, 1, 2, 3 \}$ and an orthogonal subspace containing $\gamma^{\mu} = \bar{\gamma}^{\mu}, \mu \in \{ 4, \ldots, d-1 \}$. In this case, if we define the anticommutation relation as
\begin{equation}
\{ \gamma^{\mu}, \gamma_{5} \} = \begin{cases}
\;\;\;0, & \mu \in 0, 1, 2, 3\\
2\bar{\gamma}^{\mu}\gamma_{5}, & \;\;\text{otherwise}
\end{cases}
, \quad \gamma_{5} \equiv \frac{i}{4!}\epsilon_{\mu\nu\rho\sigma}\hat{\gamma}^{\mu}\hat{\gamma}^{\nu}\hat{\gamma}^{\rho}\hat{\gamma}^{\sigma}
\end{equation}
then trace-cyclicity is consistent with \eqref{abj}. However, this method forces the fermion propagator to take its four-dimensional form for all $d$, hence fermion loops cannot be regularized; attempting to use the $d$-dimensional propagator in loop integrals is then equivalent to adding an additional term $\delta \mathscr{L} = \bar{\psi}i\bar{\gamma}^{\mu}\partial_{\mu}\psi$ to the Lagrangian density, explicitly breaking gauge invariance \cite{J00}.

Thus, if one wishes to renormalize a gauge theory with chiral fermions, one must sacrifice either cyclicity of the trace over Dirac matrices involving $\gamma_{5}$, or break gauge invariance at intermediate stages of a calculation in perturbation theory. The former option is preferable for the purpose of calculating higher-order perturbative corrections, but will inevitably give rise to ambiguities in loop integrals stemming from the precise location of $\gamma_{5}$ in the Dirac traces. Such ambiguities may appear for the first time at three loops, however the $\beta$-functions of the gauge \cite{MSS13} and scalar \cite{BPV13} couplings in the Standard Model are spared, due to the cancellation of the ABJ anomaly. Furthermore, for the Yukawa couplings, one can use a ``semi-na\"ive" treatment of $\gamma_{5}$,
\begin{equation}
\text{tr}\left[ \gamma^{\mu}\gamma^{\nu}\gamma^{\rho}\gamma^{\sigma}\gamma_{5} \right] = 4i\tilde{\epsilon}^{\mu\nu\rho\sigma} + \mathcal{O}(\epsilon), \quad\quad \tilde{\epsilon}^{\mu\nu\rho\sigma}\tilde{\epsilon}_{\alpha\beta\gamma\delta} = g^{\left[\mu\right.}_{\;\left[\alpha\right.}g^{\nu}_{\;\beta}g^{\rho}_{\;\gamma}g^{\left.\sigma\right]}_{\;\left.\delta\right]}, \quad\quad \tilde{\epsilon}^{\mu\nu\rho\sigma} \xrightarrow{d \rightarrow 4} \epsilon^{\mu\nu\rho\sigma}
\end{equation}
in order to show that the resulting ambiguity in the relevant Feynman integral is $\mathcal{O}(\epsilon)$, and hence cannot affect the Yukawa $\beta$-function \cite{CZ12}. Unfortunately, such minor miracles no longer hold at four loops; by parametrizing the integrals according to the ``reading point" of the traces, the resulting ambiguity in the four-loop strong-coupling $\beta$-function, $\beta_{a_{S}}^{(4)}$, has been explicitly calculated \cite{BP16,Z16}. In the conventions of \cite{BP16}, given rescaled couplings $(16\pi^{2})a = \left\{ g_{S}^{2}, y_{t}^{2} \right\}$ and $\beta$-functions defined by $\tfrac{\text{d} a_{S}}{\text{d}\ln \mu^{2}} = \beta_{a_{S}}a_{S}$, one finds
\begin{equation}
\beta^{(4)}_{a_{S}} \;\supset\;  R\left(\frac{16}{3} + 32\zeta_{3}\right)\,T_{F}^{2}a^{2}_{S}a_{t}^2, \quad R \overset{?}{=} 1, 2, \text{or\;} 3.
\label{ambi}
\end{equation}

While the pursuit of higher-order loop calculations has motivated many significant computational developments, there have also been notable advances in our understanding of renormalization itself, which are not yet as well-known in the phenomenological community. One such development is the notion of Weyl Consistency Conditions \cite{O90}: if one extends a theory to curved spacetime and local couplings, then the Wess-Zumino consistency conditions for the trace anomaly imply a plethora of relations between various RG quantities, amongst them Osborn's equation\footnote{This equation is not generally known by any set name, but one of us (CP) is fed up of using phrases such as the technically incorrect ``gradient-flow equation", the correct-but-cumbersome ``gradient-flow-like equation", and the frankly horrific ``equation defining the four-dimensional perturbative $a$-function". As the power of this equation is only now being realized, we feel it appropriate that its author be suitably recognised.}
\begin{equation}
\partial_{I}\tilde{A} \equiv \frac{\partial \tilde{A}}{\partial g^{I}} = T_{IJ}\beta^{J},
\label{gfe}
\end{equation}
where $g^{I}$ labels the marginal couplings of the theory. This equation therefore demonstrates the existence of a scalar function, $\tilde{A}$, of the couplings in a general renormalizable theory, which places constraints on the corresponding $\beta$-functions -- $T_{IJ}$ is simply a function of the couplings, and the form of its perturbative expansion is fixed by \eqref{gfe}. Central to these constraints is the ``3--2--1" phenomenon, where the gauge $\beta$-function is related to the Yukawa $\beta$-function one loop below, and the scalar $\beta$-function two loops below. The reason for this ordering is topological, and is thus manifestly preserved to all orders; consequently, given enough information at lower orders, one can use \eqref{gfe} to predict coefficients of terms at higher orders. Most importantly, the $\beta$-functions in \eqref{gfe} are precisely the \emph{four-dimensional} functions that one should obtain after taking the $\epsilon \rightarrow 0$ limit of Dimensional Regularization. This is the crux of our approach: if there exists a consistency condition relating the ambiguous term in $\beta_{a_{S}}^{(4)}$ to lower-order $\beta$-function coefficients, and if the consistency condition is simple enough, then it may be possible to fix the ambiguity inherent in the treatment of $\gamma_{5}$.

This paper is essentially a companion piece to \cite{WCC}, which contains a full and detailed analysis of the $\beta$-function constraints imposed by Osborn's equation, with all non-trivial modifications, up to order $\tilde{A}^{(5)}$. We begin with a summary of the Lagrangian density for a general, renormalizable, four-dimensional theory, then introduce our diagrammatic notation for the associated tensor structures appearing in \eqref{gfe}. We then quickly re-derive the constraints on $\gamma_{5}$ contributions, using a topological shortcut. Finally, we use \cite{BPV14} to reconstruct the \msbar coefficients for all terms in the general 3-loop Yukawa $\beta$-function that involve $\gamma_{5}$; the constraints then uniquely determine all $\gamma_{5}$ contributions to the general 4-loop gauge $\beta$-function, from which we extract the unique, consistent value for $R$ in \eqref{ambi}.

\section{General theory and diagrammatic notation}
In order to derive constraints on the four-loop gauge $\beta$-function, one must construct $\tilde{A}$ at five loops. This is already a somewhat awkward task, but there is a further complication: in order to isolate particular contributions to the $\beta$-function, such as those stemming from the integrals involving $\gamma_{5}$, one must work with a completely general theory, described in terms of tensor couplings between arbitrary multiplets of matter fields. The most general, renormalizable, four-dimensional Lagrangian density, based on a compact gauge group $ \mathcal{G} = \times_u \mathcal{G}_u $ with any number of Abelian and non-Abelian factors, can be written as
\begin{align}
\mathscr{L} &= -\tfrac{1}{4} \sum_{u} F_{u,\mu\nu}^{A_u} F^{A_u \mu\nu}_{u} + \tfrac{1}{2} (D_\mu \phi)_{a} (D^{\mu} \phi)_a + i \bar{\psi}_i \bar{\sigma}^\mu (D_\mu \psi)^i\nn
& \quad - \tfrac{1}{2}\left( Y^{aij}\phi_{a}\psi_{i}\psi_{j} + \bar{Y}^{a}_{ij}\phi_{a}\bar{\psi}^{i}\bar{\psi}^{j}\right) - \tfrac{1}{4!}\lambda_{abcd}\phi_{a}\phi_{b}\phi_{c}\phi_{d}\nn
& \quad + \text{\;mass terms\;} + \text{\;relevant operators\;} + \text{\;gauge-fixing\;} + \text{\;ghost terms\;}
\label{lagrangian}
\end{align}
with covariant derivatives
\begin{equation}
D_\mu \phi_a = \partial_\mu \phi_a -i\sum_u g_u \,V_{u,\mu}^{A_u}  (T_{\phi,u}^{A_u})_{ab} \phi_b, \quad D_\mu \psi^i = \partial_\mu \psi^i -i\sum_u g_u\,V_{u,\mu}^{A_u}  (T_{\psi,u}^{A_u})\udindices{i}{j} \psi^j.
\end{equation}
The fermions transform under a representation $R_{u}$ of the corresponding gauge group $\mathcal{G}_{u}$, with Hermitian generators $(T_{\psi,u}^{A_u})^{\dagger} = (T_{\psi,u}^{A_u})$; likewise, the scalars transform under a real representation $S_{u}$ with antisymmetric, Hermitian generators $(T_{\phi,u}^{A_u})^{T} = -(T_{\phi,u}^{A_u})$.

For our purposes, it proves convenient to assemble the Yukawa couplings and fermion generators into larger matrices,
\begin{align}
\quad y_a &= \begin{pmatrix}	Y_a & 0 \\ 0 & Y_{a}^{\ast }\end{pmatrix}, & T^{A_u}_u &= \begin{pmatrix} T_{\psi,u}^{A_u} & 0 \\ 0 & -(T_{\psi,u}^{A_u})^\ast \end{pmatrix},  \nn
\tilde{y}_a = \sigma_1 y_a \sigma_1 &= \begin{pmatrix}
Y_a^{\ast} &0 \\ 0 & Y_a
\end{pmatrix}, & \tilde{T}^{A} &= \sigma_1 T^{A} \sigma_1 = - (T^{A})^{T},
\label{Yukawa}
\end{align}
so that there is a single Yukawa interaction between Majorana-like spinors $ \Psi_i = \binom{\psi }{\bar{\psi} } $. Similarly, by arranging $\mathcal{G}$ such that the first $n$ factors are Abelian, we may define gauge field multiplets $A^{A}_{\mu}$ with a generalized adjoint index $A$:
\begin{equation}
A \in \{(u, A_u): \; A_u \leq d(\mathcal{G}_u) \} \quad \text{with summation convention} \quad \sum_{A} = \sum_{u} \sum_{A_u =1}^{d(\mathcal{G}_u)}.
\end{equation}
The gauge couplings may then be assembled into a block-diagonal matrix,
\begin{equation}
G_{AB}^2 = \begin{cases}h_{uv}^2 & \text{for } A, B\leq n \\
g_u^2 \delta_{uv} \delta^{A_u B_v} & \text{for } A > n \end{cases},
\end{equation}
where $h^{2}_{uv}$ is a symmetric $n\times n$ matrix of $U(1)$ couplings, allowing us to incorporate the effects of kinetic mixing\footnote{The effect of kinetic mixing on the $\beta$-functions was first derived in \cite{LX03}, and was reinterpreted in terms of a unified matrix coupling in \cite{FMS14}.}; generalized group Casimirs $[C_{2}(G)]_{AB}$ and Dynkin indices $[S_{2}(R)]_{AB}, R \in \{ F, S \}$ may defined analogously. The general Lagrangian density \eqref{lagrangian} may now be re-written in the form
\begin{align}
\mathcal{L} = &-\tfrac{1}{4} G^{-2}_{AB} F^{A}_{\mu\nu} F^{B\mu\nu} + \tfrac{1}{2} (D_\mu \phi)_a (D^\mu \phi)_a + \tfrac{i}{2} \Psi^{T} \begin{pmatrix} 0 & \sigma^\mu \\ \bar{\sigma}^\mu & 0 \end{pmatrix} D_\mu \Psi \nn
&- \tfrac{1}{2} \phi_{a}\, \Psi^{T}  y_{a} \Psi -\tfrac{1}{24} \lambda_{abcd} \phi_a \phi_b \phi_c \phi_d + \ldots
\label{general}
\end{align}
with general tensor couplings $g^{I} = \{ G^{2}_{AB}, y_{aij}, \lambda_{abcd} \}$, and the associated $\beta$-functions defined according to $\beta^{I} = \tfrac{ \text{d} g^{I}}{\text{d} \ln \mu}$.

When evaluating loop integrals, the Feynman rules for each diagram produce contractions between the various tensor couplings and group factors, which we refer to as Tensor Structures (TSs); the $\beta$-functions may therefore be expressed as a sum over particular TSs, each weighted by a coefficient\footnote{Scheme-dependence of the $\beta$-functions then simply corresponds to changes in these coefficients.}. To maintain gauge invariance of a theory with multiple interactions, the generators and tensor couplings must satisfy
\begin{align}
0 &= -\tilde{T}^{A} y_a + y_a T^{A} + y_b (T_{\phi}^{A})_{ba},\nn
0 &= (T^{A}_\phi)_{ae} \lambda_{ebcd} +(T^{A}_\phi)_{be} \lambda_{aecd} + (T^{A}_\phi)_{ce} \lambda_{abed} +(T^{A}_\phi)_{ad} \lambda_{abce}.
\label{gauge}
\end{align}
Thus, there will be non-trivial relations between TSs, and one must reduce the full set of TSs in each $\beta$-function to a basis. Once done, the TSs may be represented using a convenient diagrammatic notation, based on the following identifications:
\begin{equation}
\begin{gathered}
\vcenter{\hbox{\includegraphics{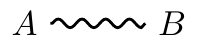}}} = G^2_{AB}, \qquad 
\vcenter{\hbox{\includegraphics{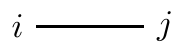}}} = \delta_{ij}, \qquad
\vcenter{\hbox{\includegraphics{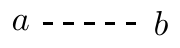}}} = \delta_{ab},
\\
\vcenter{\hbox{\includegraphics{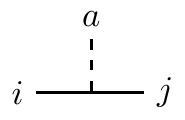}}} = y_{aij}, \qquad
\vcenter{\hbox{\includegraphics{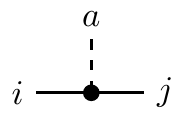}}} = (\sigma_3 y_{a})_{ij}, \qquad
\vcenter{\hbox{\includegraphics{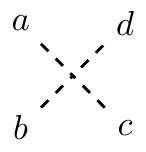}}} = \lambda_{abcd}, \\
\vcenter{\hbox{\includegraphics{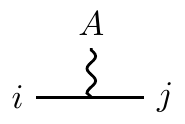}}} = (T^{A})_{ij}, \qquad \vcenter{\hbox{\includegraphics{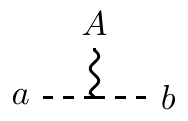}}} = (T^{A}_\phi)_{ab}, \qquad
\vcenter{\hbox{\includegraphics{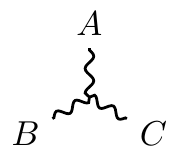}}} = G^{-2}_{AD} f^{DBC}.
\end{gathered}
\label{diagrams}
\end{equation}
Contracted indices in TSs (with summation implied) are then represented by lines connecting different vertices. Since the Feynman rules for \eqref{general} introduce a $\sigma_{1}$ between two Yukawa couplings, TSs implicitly alternate between $y_{a}$ and $\tilde{y}_{a}$ along a fermion line. Note that the assembled Yukawa matrices in \eqref{Yukawa} automatically incorporate contributions from left- and right-handed fermions with the same overall sign, whereas $\gamma_{5}$ contributions give the \emph{opposite} sign. To accommodate this, we simply insert a  $\sigma_{3}$($\tilde{\sigma}_{3}$) factor along with $y_{a}$($\tilde{y}_{a}$), representing such insertions with a blob on the Yukawa vertex.

The two main advantage of this representation are that it becomes substantially easier to represent and visualize $\beta$-functions at higher loop-orders, and that the notation may be extended to all other quantities appearing in \eqref{gfe}. A simplified version of this notation was used in \cite{JP14} to construct $\tilde{A}$ at four loops for a theory with a simple gauge group, and to derive the associated consistency conditions. Taking all this into account, the $\beta$-functions of the general theory described by \eqref{general} may be expanded in the form
\begin{equation} \label{eq:beta_parametrization}
\beta_{AB} = \dfrac{\dd G^{2}_{AB}}{\dd \ln\mu} = \dfrac{1}{2} \sum_{\mathrm{perm}} \sum_{\ell} G^{2}_{AC} \dfrac{ \beta^{(\ell)}_{CD} }{(4 \pi)^{2\ell}} G^{2}_{DB}, \qquad
\beta_{aij} = \dfrac{\dd y_{aij}}{\dd \ln \mu} = \dfrac{1}{2} \sum_{\mathrm{perm}} \sum_{\ell} \dfrac{\beta^{(\ell)}_{aij}}{(4\pi)^{2\ell}},
\end{equation}
with associated $\ell$-loop diagrammatic representations given by
\begin{equation}
\beta^{(\ell)}_{AB} = \sum_n \cofg{\ell}{n} \!\! \vcenter{\hbox{\includegraphics{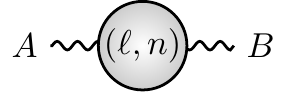}}} , \qquad
\beta^{(\ell)}_{aij} = \sum_n \cofy{\ell}{n} \!\! \vcenter{\hbox{\includegraphics{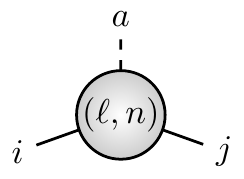}}}.
\end{equation}
The blobs represent a suitable basis of TSs with the correct external legs, constructed using the identifications in \eqref{diagrams}, each multiplied by a coefficient.

\section{Constraints from Weyl Consistency Conditions}
\begin{figure}[t]
	\centering
	\includegraphics[width=\textwidth]{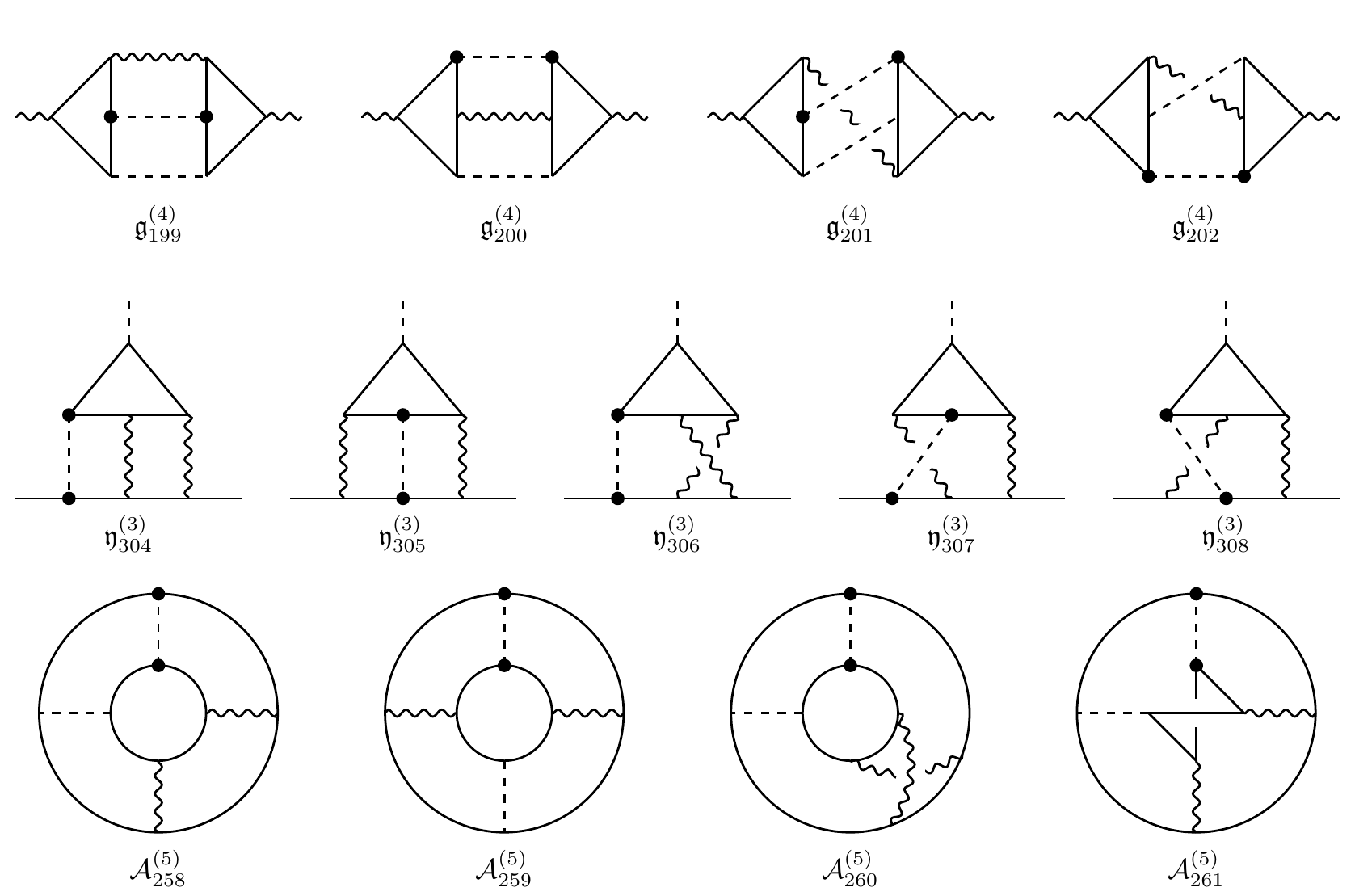}
	\caption{Complete set of Tensor Structures (TSs) related to the first non-trivial contributions involving $\gamma_{5}$, taken from \cite{WCC}. The first two rows are contributions to $ \beta^{(4)}_{AB} $ and $ \beta^{(3)}_{aij} $ respectively, while the last row gives the 4 corresponding TSs in $ \tilde{A}^{(5)} $. The blobs on Yukawa vertices symbolize $ \sigma_3 $ insertions.}
	\label{fig}
\end{figure}

To construct $\tilde{A}$ in Osborn's equation, it is more convenient to work with an equivalent statement, obtained by multiplying both sides of \eqref{gfe} by $\text{d}g^{I}$:
\begin{equation}
\text{d}\tilde{A} \equiv \text{d}g^{I}\partial_{I}\tilde{A} = \text{d}g^{I}T_{IJ}\beta^{J}.
\label{gfe2}
\end{equation}
Since $\tilde{A}$ is a scalar function of the couplings, its contributions may be represented by totally contracted TSs, and its total derivative should consist of contractions between the differentials $\text{d}g^{I}$ and the $\beta$-functions $\beta^{J}$. The $\ell$-loop contributions to $\tilde{A}$ are therefore given by all possible $\ell_{1}$-loop contractions between $\text{d}g^{I}$ and the $\ell_{2}$-loop $\beta^{J}$, such that $\ell = \ell_{1} + \ell_{2}$; each possible contraction represents a term in the tensor $T_{IJ}$. At leading order in each tensor coupling, and using the same conventions as \cite{WCC}, we have
\begin{equation}
T^{(1)}_{GG} = \cofT{1}{gg} G^{-2}_{AC}G^{-2}_{BD}, \qquad\qquad T^{(2)}_{yy} = \cofT{2}{yy} \delta_{ab}\delta_{ik}\delta_{jl}.
\end{equation}
To identify $\beta$-function TSs from $\gamma_{5}$ contributions, one need only consider the possible Lorentz indices that could lead to two fully contracted $\epsilon$-tensors. Of all possible diagrams at this order, only those with two fermion lines containing at least two gauge generators and one Yukawa tensor will contribute: all other possibilities vanish by anomaly cancellation. A convenient basis for the relevant diagrams occurring in $\beta^{(3)}_{aij}$ and $\beta^{(4)}_{AB}$, plus the associated terms in $\tilde{A}^{(5)}$, is given in Fig.~\ref{fig}.

It is at this point that certain special features of these terms become obvious. In principle, $\tilde{A}^{(5)}$ receives additional contributions from inserting lower-loop $\beta$-functions into higher-loop $T_{IJ}$ contractions. However, the contributions $\tilde{A}^{(5)}_{258-261}$ are topologically equivalent to cubes and M\"obius-ladders: both graphs are simple and vertex-transitive, thus the only way each tensor could receive additional contributions would be if the $\beta$-function graphs contained subgraphs that also appeared at lower loops. As each $\beta$-function graph is primitive, this is clearly not the case, hence there are no other possible contributions to these tensors. Substituting the terms from Fig.~\ref{fig} into \eqref{gfe} therefore gives
\begin{align}
2 \cofA{5}{258} & = T^{(1)}_{gg,1} \cofg{4}{199}, & 2 \cofA{5}{259} & = T^{(1)}_{gg,1} \cofg{4}{200}, & 2 \cofA{5}{260} & = T^{(1)}_{gg,1} \cofg{4}{201}, & 2 \cofA{5}{261} & = T^{(1)}_{gg,1} \cofg{4}{202}, \nn
4 \cofA{5}{258} &= T^{(1)}_{yy,1} \cofy{3}{304}, & 4 \cofA{5}{259} &= T^{(1)}_{yy,1} \cofy{3}{305}, &  4 \cofA{5}{260} &= T^{(1)}_{yy,1} \cofy{3}{306}, &  2 \cofA{5}{261} &= T^{(1)}_{yy,1} \cofy{3}{307}, \\
&& && && 2 \cofA{5}{261} &= T^{(1)}_{yy,1} \cofy{3}{308}. \nonumber  
\end{align}
Eliminating the $\tilde{A}^{(5)}$ coefficients and substituting in the (scheme-independent) $T_{IJ}$ coefficients \cite{JO91} then leaves five consistency conditions, completely determining the $\gamma_{5}$ contributions to $\beta^{(4)}_{AB}$:
\begin{equation}
\cofg{4}{199} = \dfrac{1}{6}\cofy{3}{304}, \quad 	
\cofg{4}{200} = \dfrac{1}{6}\cofy{3}{305}, \quad
\cofg{4}{201} = \dfrac{1}{6}\cofy{3}{306}, \quad
\cofg{4}{202} = \dfrac{1}{3}\cofy{3}{307}, \quad
\cofy{3}{307} = \cofy{3}{308}.
\label{cc}
\end{equation}

\section{Standard Model $\beta$-functions and $\gamma_{5}$}
The consistency conditions \eqref{cc} relate TSs that may receive non-trivial contributions from integrals involving $\gamma_{5}$, and hold for a completely general renormalizable theory with a compact gauge group. The Standard Model is, of course, precisely such a theory, so we may deduce $\cofy{3}{304-308}$ by substituting the SM couplings into a basis of relevant $\beta^{(3)}_{aij}$ TSs\footnote{The $\gamma_{5}$ contributions to $\beta^{(3)}_{aij}$ are listed in Fig.~\ref{fig}, and the full basis of $\beta^{(3)}_{aij}$ TSs is given in the ancillary file of \cite{WCC}.}, and comparing with the known SM results of \cite{BPV14}. It is sufficient to focus on TSs containing four powers of the gauge couplings and a trace over two Yukawa matrices; the comparison may be done using either Appendix D of \cite{MSS13}, or the general methods of \cite{M14}. Our final result is
\begin{equation}
\cofy{3}{304} = -24, \qquad \cofy{3}{305} = -12, \qquad \cofy{3}{306} =  \cofy{3}{307} =  \cofy{3}{308} = 8 - 24\zeta_{3},
\label{ysol}
\end{equation}
hence equation \eqref{cc} requires that
\begin{equation}
\cofg{4}{199} = -4, \qquad \cofg{4}{200} = -2, \qquad \cofg{4}{201} = \tfrac43 - 4\zeta_{3}, \qquad \cofg{4}{202} = \tfrac83 - 8\zeta_{3},
\label{gsol}
\end{equation}
as well as confirming that the final condition, $\cofy{3}{307} = \cofy{3}{308}$, is indeed satisfied. Substituting the SM couplings into the $\beta^{(4)}_{AB}$ TSs, multiplying by the coefficients in \eqref{gsol}, and converting to the conventions of \cite{BP16} then gives
\begin{equation}
\beta^{(4)}_{a_{S}} \;\supset\; (16 + 96\zeta_{3})\,T_{F}^{2}a^{2}_{S}a_{t}^2.
\end{equation}
Thus, comparison with \eqref{ambi} forces the choice
\begin{equation}
R = 3
\end{equation}
in the $\beta_{a_{S}}^{(4)}$ calculation of \cite{BP16, Z16}, corresponding to a reading whereby one cuts the traces at any of the internal vertices. While \cite{BP16} gave some theoretical justifications for preferring this value of $R$, we believe this constitutes the first proof that it must be so; furthermore, our results determine \emph{all} $\gamma_{5}$ contributions to \emph{any} gauge $\beta$-function at four loops, including all three SM gauge couplings with full matter content. We stress that there is no wiggle-room in the conclusion: \eqref{cc} relates the final $\beta$-function coefficients after removal of the regulator, and holds for all perturbative renormalization schemes, thus the four-loop integral involving $\gamma_{5}$ \emph{must} be treated in this manner.\\

The topological argument guaranteeing that no higher-order $T_{IJ}$ contributions influence the consistency conditions can easily be extended to higher loops: if the tensor structure in $\tilde{A}^{(n)}$ is topologically equivalent to a connected symmetric graph\footnote{A symmetric graph generally refers to a graph with a set number of edges connected to each vertex, such that the automorphism group acts transitively on both the associated vertex- and edge-graph; a connected symmetric graph is then a symmetric graph with no disconnected vertices or subgraphs. Due to the multiple interaction types, the graph topologies that contribute to the $A$-function and lead to a simple consistency condition like \eqref{cc} are more general - we are unaware of a classification scheme for all such topologies, but the connected symmetric graphs form a well-defined subset.}, and the associated primitive tensors in $\beta_{AB}^{(n-1)}$, $\beta_{aij}^{(n-2)}$ and/or $\beta_{abcd}^{(n-3)}$ contain non-trivial contributions from $\gamma_{5}$, then one can quickly derive an analogous consistency condition to fix the potential ambiguity, as parametrized by the same trace-cutting procedure used at four loops. It may of course be possible that, at higher orders, $\gamma_{5}$ contributions also appear in more complex conditions than those similar to \eqref{cc}. If this is so, it is still possible to use the full set of consistency conditions to infer a consistent treatment, although the amount of work required will be dramatically increased.

\subsection*{Acknowledgements}
We are very grateful to Florian Herren for useful discussions, helping clarify the treatment of $\gamma_{5}$ in general theories. CP would like to thank Joshua Davies for his off-the-cuff question that eventually led to the result in this paper, and Ian Jack for his careful reading of the manuscript. AET would like to thank Fermi National Accelerator Lab for hosting him during the completion of this paper, and gratefully acknowledges financial support from the Danish Ministry of Higher Education and Science through an EliteForsk Travel Grant. This work is partially supported by the Danish National Research Foundation grant DNRF90.

\end{document}